\documentclass[a4paper,11pt]{article}

\usepackage{graphicx}
\usepackage{dcolumn}% Align table columns on decimal point
\newcommand{\beq}{\begin{equation}}
\newcommand{\eeq}{\end{equation}}

\title{The dynamics of spontaneous emission}
\author{M. Baldo}
\date{ }

\begin{document}
\maketitle
\vspace {-1 cm}
\begin{center}
\textit{Istituto Nazionale di Fisica Nucleare}
\par \textit{via Santa Sofia 64, 95123 Catania, Italy} 
\end{center}
\vskip 0.5 cm
\noindent
Abstract
\par
The spontaneous decay of an excited atom by photon emission is one of the most common and elementary physical process present in nature and in laboratories. The decay is random in time with constant probability density, as it can be inferred by the exponential law observed experimentally. If a large set of independent and isolated atoms are all equally excited at a given time, the number of atoms still excited after a time $ t $ decreases by a factor $ \exp (-\Gamma t) $, where $ \Gamma $ is the decay rate. Despite the simplicity of the process, in Quantum Mechanics the decay itself is considered a law of nature which is not further analyzed or "explained". However it is legitimate to ask for the reason of its randomness and for the dynamics of the atom around the decay. The decay process of an isolated atom is usually assumed to be instantaneous, the so called "Quantum jumps". Particular experimental arrangements can widen the duration of the transition from an excited state to the ground state to a finite time. In general this is due to the quantum back     
action of the detector. The development of Quantum Optics has enormously enriched the possibilities to study in detail the process, and new illuminating insights have been obtained, but the question is still valid and its answer is still elusive.
In this paper we analyze the spontaneous decay within the new prospect introduced by a recent model that complete standard Quantum Mechanics in a formalism which is able to describe the dynamics of measurement and of spontaneous decay. Comparison is made with more phenomenological theories, the "Quantum Mechanics of open system" and the "Stochastic wave function" model. 
In the new model the stochastic photon emission by an excited atom is triggered by the vacuum fluctuations. It is suggested how to re-analyze three types of experiments, already realized in many laboratories, in order to reveal the presence and the effect of these fluctuations. 

\section{Introduction}
\par 
Since the early beginning of Quantum Mechanics (QM) it was assumed that an isolated excited atom will undergo a spontaneous decay 
to a lower energy level by emitting a photon \cite{first}, which was the basis for the explanation of atomic spectra. 
The emission is considered random, with a uniform probability density in time. These assumptions are still valid today, 
within the full development of QM. The characteristic of the spontaneous emission is therefore twofold
\par\noindent
1. The photon emission is indeed "spontaneous", i.e. it can occur without need of an external perturbation.
\par\noindent
2. The emission occurs randomly with constant probability density in time.
\par 
These two assumptions explain the observed "exponential law". One finds experimentally that the probability for an isolated excited atom to not decay up to
the time $t$ decreases exponentially as $\exp( -\Gamma t )$, where $ \Gamma $ is the decay constant for a given
decay process, and in fact it is the decay probability density per unit of time.  \par 
These two postulates pose some natural questions. First of all the term itself "spontaneous" sounds a little too
anthropomorphic, suggesting that the atom has the ability to choose the time of decay. Apart this "linguistic" remark, it appears legitimate to ask for the origin of the randomness of the decay and of its uniform probability. Finally, 
the decay itself is abrupt, i.e. a "quantum jump", usually assumed instantaneous \cite{jump1,jump2,jump3}, and therefore one can ask what is the dynamics of the atomic wave function in the proximity of the jump. \par 
If we follow the ordinary QM formalism, it must be noticed that the initial excited state cannot be a sharp energy
eigenstate. In fact each atomic excited state has a width, the so called "natural width", just because it is coupled to the continuum of the photon field, i.e. it is a "resonance" in 
broad sense. Therefore the initial state evolves by developing two components, one corresponding to the excited state and one corresponding to the ground state plus a photon (for simplicity we consider only these two levels). The strengths of the two components change with time, with the excited state component decreasing in time and the other increasing in time, keeping the overall normalization constant. At a certain time the decay occurs, and only the ground sate component remains and an emitted photon. If a detector is present, it can absorb the photon before the decay occurs, and the ground state component is then selected by the measurement process, which implies a quantum back action, because of the coherence between the two states before the detection process. All that can be described by keeping the unitary evolution of the atomic state, supplemented by the random decay process, which evidently is not unitary. 
\par 
Several theories present in the literature provide different descriptions of the wave function evolution around a spontaneous decay. Most of them take the quantum jump as a fundamental process that cannot be further analyzed,
i.e. it is granted by nature without an underlying origin. However they differ in the dynamics, especially if the decay is occurring in more complex physical conditions, in particular in case that the atom is subject to a forced Rabi oscillation. In general the dynamics in these theories is different from the unitary one because the atomic system is
considered open, and therefore under the influence of the environment. The action of the environment can be modeled and  interpreted in different ways, which in general induces a different dynamics. \par
The problem of the decay quantum jump and of its randomness was considered within a realistic extension of QM \cite{IJQF}, which, while keeping the whole results and structure of the standard QM, introduces new degrees of freedom that are ruling the wave function reduction process, as present in e.g. a strong measurement, and the corresponding Born' s law. The spontaneous decay is then viewed as a reduction process produced by the vacuum fluctuations, and it turns out that the wave function evolution even before the decay is not strictly unitary.\par 
All that applies straightforwardly to a general quantum system that is equivalent to a single qubit, like some superconducting devises \cite{Houck}. \par
It is the purpose of this work to explain in detail the theory suggested in ref. \cite{IJQF} and to compare it critically with more phenomenological theories on the decay dynamics, and emphasize their similarities and differences. Furthermore it will be suggested some possibilities to reveal experimentally the vacuum fluctuations. 
\par
\section{A single isolated atom. \label{single}}
\par
In this section we discuss the simplest case of the decay of a single atom. Despite it is the most elementary process, it can shade light on the dynamics of the photon emission.
It is convenient to specify as clear as possible the physical conditions where the spontaneous decay of an atom, or one analogous qubit, actually occurs. As mentioned in the introduction, standard quantum mechanics predicts that an excited atomic state has a natural width $\Gamma$. This means that the starting resonance is not a stationary state, but it evolves by developing a component in the continuum, which will be the product sate of the atom ground state and of
 a photon wave packet. The latter has a width $ c \Gamma $ in momentum, and therefore it is of size $ \emph{l} \,=\,\hbar/c\Gamma \,=\, c \tau $ in space, where $ \tau $ is the lifetime of the excited state. It follows that the size $ \emph{l} $ can be of the order of millimeters for the shorter lifetimes up to several kilometers for the longer lifetimes, or even larger for metastable excited states. More formally, the evolution of the initial excited state $ | \Psi_1(0) > $ can be specified as

\begin{eqnarray}
U(t) | \Psi_1(0) > &\,=\,& \, C_1(t) | \Psi_1(0) > \,+\,\int d E_\gamma C_{0\gamma}(t) | \Psi_0(0) > | E_\gamma > \nonumber \\
%  \ &\ \nonumber \\
 | C_1(t) |^2 &\,=\,& \exp(-\Gamma t) \nonumber \\
%  \ &\  \nonumber \\
 \int d E_\gamma | C_{0\gamma}(t) |^2  &\,=\,& 1 \,-\, \exp(-\Gamma t) 
 \label{eq:U} \end{eqnarray}
\par\noindent
where $ U(t) $ is the unitary evolution operator and $ | E_\gamma > $ the photon plane wave of energy $ E_\gamma $ (for simplicity we omit polarization).
This evolution can be derived formally \cite{IJQF} and it is valid to all orders in the coupling between the excited state and the photon field. However it is valid only after the short transient period where the evolution is quadratic in time,
which we will neglect in the following.
This evolution must be implemented by the random quantum jump process that can occurs at any time with constant probability density $ \Gamma $. Of course once the jumps occurs, the atom will remain in the ground state. Let us first disregard the jumps and let us suppose that a $ 4\pi $ detector with $ 100\% $ efficiency is placed at distance $ L $ from the atom (and nothing else). If $ L >> {\it l} $ the detection of the emitted photon will always occur at fixed time $ t \,=\, L/c $, and obviously no exponential law can appear. This is equivalent to assume that the atom decay process occurs at the moment of detection. This shows the need of the quantum jump, which occurs without the action of the detector or of any other external perturbation. In fact, if we divide a generic time $ t $ in $ n $ small enough interval $ \Delta t \,=\, t/n $,
the probability $ P(t) $ that the atom has not decayed up to $ t $ will be
\beq
P(t) \,=\, (1 - \Gamma \Delta t)^n = (1 -\Gamma t/n)^n\, \rightarrow\, \exp( -\Gamma t )
\label{eq:exp}\eeq
\noindent
Notice that the decay probability $ \Gamma \Delta t $ in a generic interval of size $ \Delta t $ strictly speaking cannot be derived by first order perturbation theory, which relies on the large time limit \cite{Mes}, while $ \Delta t $ in Eq. (\ref{eq:exp}) has to be taken as small as possible.\par 
The decay dynamics in the formalism of "quantum mechanics of open systems" (QMOP) \cite{Carmib} the continuum component is treated as a dissipative part of the evolution, which therefore is not unitary.  If the initial state of the atom is more generally a linear combination of the excited state and the ground state
\beq
\Psi(0) \,=\, C_1(0) \Psi_1 \,+\, C_0(0) \Psi_0
\label{eq:init}\eeq
\noindent
the equations of motion for the coefficients read \cite{Carmib} 
\begin{eqnarray}
\frac{d}{dt} C_1(t) &\,=\,& - (\, \Gamma/2 + \imath\omega_1\, )\, C_1(t) \nonumber \\
\frac{d}{dt} C_0(t) &\,=\,& - \imath\omega_0\, C_0(t)
\label{eq:OPeq}\end{eqnarray}
\noindent
where the energies of the two states have been introduced. The corresponding Hamiltonian is not hermitian and therefore it is an effective one, where the photon field does not appear explicitly.
The first of the Eqs. (\ref{eq:OPeq}) introduces a width $ \Gamma $ for the excited state, which then will decay, while the second equation implies that the ground state component will remain of constant modulus. Because of the non unitarity of these equations, the norm of the state is not conserved. It has to be normalized, and the final expressions for the coefficients are
\begin{eqnarray}
C_1(t) &\,=\,& C_1(0)\exp(-\Gamma t/2 -\imath \omega_1 t)/N \nonumber \\
C_0(t) &\,=\,& C_0(0)\exp(-\imath \omega_0 t)/N \nonumber \\
N &\,=\,& [\, |C_0(0)|^2 + |C_1(0)|^2 \exp(-\Gamma t)\, ]^{\frac{1}{2}}   
\label{eq:norm}\end{eqnarray}
\noindent
The probability $ P_{\Delta t}(t) $ of decay in the short interval $ (t, t+\Delta t) $ is   
\beq
 P_{\Delta t}(t) \,=\, |C_1(t)|^2 \Gamma \Delta t \,=\, \Gamma \Delta t |C_1(0)|^2 \exp( -\Gamma t )/N
\label{eq:conpro}
\eeq
\noindent
This expression is interpreted as the conditional probability of decay once no decay has occurred up to $ t $, which
gives the meaning of the normalization procedure. 
If $ C_0(0) = 0 $, i.e. the initial state is the excited state, the probability of Eq. (\ref{eq:conpro}) reduces to
$ \Gamma \Delta t $, which is independent of time, and we then find again the exponential law, as shown in 
Eq. (\ref{eq:exp}).
\par 
Another approach to the spontaneous decay dynamics is the "stochastic wave function" model (SWF) \cite{SWF,Carmib}. At statistical level it provides the same master equation of QMOP, which is the basic equation for Quantum Optics, but the evolution of a single decay process turns out to be partly different. In this theory the photon field appears explicitly.
Let us assume that at time $ t $ the wave function is given by Eq. (\ref{eq:init}). After a time $ \Delta t $ the wave function acquires a component which includes a photon as in Eq. (\ref{eq:U}), with a wave packet $ C_{0\gamma} $, while the amplitude of the excited state is reduced by a proper factor to conserve the norm.
%, i.e. $ (1 - |C_{0\gamma}|^2)^{\frac{1}{2}} $. 
The amplitude is calculated in first order perturbation theory in the coupling between the electron and the photon field \cite{SWF}. At the time $ t + \Delta t $ a gedanken measurement process is performed with a hypothetical perfect detector, which is able to count the number of photons, including the case of zero photons. If $ \Delta t $ is small enough, at most one photon can be present. The probability of detecting a photon will be $ P_\gamma = \int |C_{0\gamma}(t+\Delta t)|^2 $ and of not detecting 
$ 1 - P_\gamma $. Notice that the wave function at time $ t+\Delta t$ has been normalized because of the Born' s rule of Quantum Mechanics formalism. After the measurement the atom will be either in the ground state or in the modified component with no photon. In the first case the evolution stops, while in the second case the procedure can be repeated. Here the randomness of the emission is introduced
by the gedanken measuring process. If we could take the limit of vanishing small $ \Delta t $, the dynamical evolution would coincide with the one of QMOP, and in fact the formalism is often called "continuous recording" model.
However, as already noticed, the usual expression in perturbation theory for the transition probability demands the infinite ( or large ) time limit. For a finite $ \Delta t $ the evolution of the component of the excited state would be a series of small jumps, until the final quantum jump. This implies a deviation, eventually small, from the exponential law.
Instead of the perturbation theory one can use the evolution of Eq. (\ref{eq:U}), and the exponential law can be recovered. In fact, at a given step the excited state strength would be reduced by a factor $ \exp( -\Gamma\Delta t ) $, 
and after $ n $ step of no photon recording by a factor $ \exp( -\Gamma n\Delta t ) $. In this case, however, for a sizable $ \Delta t $, the quantum optics master equation had to be modified. In the limit of vanishing $ \Delta t $, the so called "continuous" limit, the occupation number of the initial pure excited state $ \Psi_1 $ will remain equal to $ 1 $ all the time before the quantum jump, in agreement with the QMOP model.    \par 
The basic criticism to the model is on the introduction of the gedanken measurement process, where the no photon component is also recorded. To justify the procedure one usually invokes the possibility of reduction of the wave function with no interaction, the so called null experiment, where the negative result for a detection produces a modification of the wave function. The main reference for this possibility is the experiment proposed in \cite{EV,V}, where in a 
Max-Zender interferometer a perfect detector is introduced in one of the two branches. With probability $ 1/4 $ one can observe that the photon has followed the branch free of obstacle and arrived at the right port, but no interaction had happened at the detector. The original two components photon wave packet is then reduced to a single one with no detection. However this arrangement does not apply to the SWF model. In fact, the reduction of the photon wave function
with no interaction in the Max-Zender interferometer can occur only because the photon can escape from the obstacle, since it does not cover the whole area where the photon can pass, but only 50\% (for a balanced splitter). In other words, an undetected  photon is one of the possible output of the reduction process. In the SWF model it is assumed that the reduction
can take place even if the detector is able to absorb all possible photons that are present during the evolution. 
This is possible because the gedanken detector is sensitive to the zero photon state, which is difficult to justify.
Finally it is unclear what is the meaning of the presence of the hypothetical detector. 
\par 
In ref. \cite{IJQF} it was proposed a model (NSM) for the completion of QM, which is able to describe the physical processes that are kept outside the standard formalism, in particular the strong measurement process. The basis
of the model is the introduction for each wave function of a nonstandard component, in the sense of nonstandard analysis,
and of a nonstandard space-time lattice which is the support of the wave function. Under a set of assumptions, that are  coherent with the basic one, it turns out that a stochastic process must be present in the standard sector. This stochastic dynamics is in addition to the standard linear evolution ruled by Schr\"odinger equation and it can trigger the reduction of the wave function, which follows Born's rule, a process outside the formalism of standard QM. The model put the spontaneous decay in a new prospect. If an isolated atom is excited at time $ t = 0$, it starts to follow the unitary evolution of Eqs. (\ref{eq:U}). At a certain time a vacuum fluctuation occurs. The nonstandard component of the fluctuation introduces nonstandard 
frequencies in the wave function of the electrons, which in turn produces the reduction of its wave function in one of the two components, i.e. either the excited state $ \Psi_1(0) $ or the ground state $ \Psi_0(0) $ plus the wave packet
$ C_{0\gamma} $, with the probabilities dictated by the Born' s rule, as specified in the equations. In the first case the evolution starts to repeat, in the second case the atom decays in the ground state, while the photon is eventually detected. If $ \Delta t_i $ are the intervals between two subsequent vacuum fluctuations, the probability that the atom remains in the excited state will be
\beq
P(t) \,=\, e^{-\Gamma\Delta t_1}  e^{-\Gamma\Delta t_2} \cdots\dots  e^{-\Gamma\Delta t_n} \,=\,  e^{-\Gamma t}
\label{eq:exp1}\eeq   
\noindent with $ t \,=\, \sum_i \Delta t_i $. Notice that the exponential law is regained \emph{independently} from the presence of a detector, since it is the action of the vacuum fluctuations that triggers the reduction. Therefore,
on one hand there is no need of a "gedanken detector", on the other hand the origin of the "spontaneous" decay and of its randomness find an explanation. It is essential to realize that it is the nonstandard component of the vacuum fluctuations that produces the reduction of the atomic wave function. The corresponding standard component of the vacuum fluctuations, or of any other standard perturbation, cannot produce the reduction, since the evolution ruled by the Schr\"odinger equation 
cannot include any quantum jump. This is apparent from the observation of two effects of the standard component of the vacuum fluctuations, i.e. the Lamb' s shift and the Casimir effect, where clearly no reduction is involved. As discussed in ref. 
\cite{IJQF}, the duration $\Delta t$ of each fluctuation is expected to be of the order of $ \hbar/2m $, with $ m $ the electron mass. This is much shorter than any atomic electromagnetic process, which means that if the reduction results in no photon emission, the standard evolution can immediately restart with no appreciable deviation from the exponential decay. 
More uncertain is the frequency of the vacuum fluctuations, but according to Eq. (\ref{eq:exp1}), the exponential law is independent from the particular sequence of vacuum fluctuations. If the sequence of fluctuations is very rapid, or even continuous, the occupation number of the excited initial state will remain close to $ 1 $, in agreement with QMOP and SWF.
However for a discrete sequence, the occupation number between two fluctuations would be lower than $ 1 $, until the final   
quantum jump. If we neglect the duration of the fluctuations, which is indeed quite small, the occupation number is in general $ 1 - \exp(-\Gamma |\tau|) = 1 - a $, where $ |\tau| $ is the time elapsed from the last fluctuation that has not produced the quantum jump ("negative result"). The vacuum fluctuations are expected to have a constant probability to occur in an interval $ \Delta t $,
 namely $ \beta \Delta t $, with $ \beta $ the probability per unit of time. The probability density $ D_f $ that a vacuum fluctuation has not occurred in the time interval $ (\tau,0) $  is then
\beq
D_f(\tau) \,=\, \beta \exp(\beta \tau) 
\label{eq:Df}\eeq  
\noindent where $ -\infty < \tau < 0 $, and we extended the lower limit to $ -\infty $ with a negligible error if the time  
elapsed from the beginning of the evolution is long enough. This probability density can be translated in the probability density $ D(a) $ for the decrease $ a $ of the occupation number from 1 at the moment of the jump. A straightforward calculation gives ($0 < a < 1$)
\beq
D(a) \,=\, r ( 1 - a )^{r-1}
\label{eq:Da}\eeq
\noindent with $ r = \beta/\Gamma $. From this one gets the average value $ \overline{a} $ and variance $ \Delta a$ of $ a $.
One gets
\begin{eqnarray}
\overline{a} &\,=\,& \frac{1}{1+r} \,=\, \frac{\Gamma}{\beta+\Gamma}  \nonumber \\
\Delta a &\,=\,&  \frac{1}{1+r} \Big[ \frac{r}{2+r} \Big]^{\frac{1}{2}}
\label{eq:aa}\end{eqnarray}
\noindent The limit $ \beta \rightarrow \infty $ corresponds to a continuous vacuum fluctuation, and one gets 
$ \overline{a} \rightarrow 0 $ and $ \Delta a \rightarrow 0 $, which means that before the quantum jump the occupation number of the excited state remain equal to $ 1 $, in agreement with QMOP and SWF, where one assumes a continuous photon detection with negative result. In the opposite limit $ \beta \rightarrow 0 $ no vacuum fluctuation is present and the system evolves toward the ground state without any jump where of course it stays. Although it is a ill defined limit since then 
$ D_f \rightarrow 0 $, the value of 
$\overline{a} = 1$ is compatible with the ground state indefinitely occupied. This limit does not exist in QMOP and SWF, since it would correspond to the absence of any detection. In general the occupation number of the excited state will fluctuate before the quantum jump, and in principle this could be experimentally revealed by quantum tomography \cite{Brannan}, or by fluorescence emission \cite{fluo}. Therefore even in this simplest case of an excited atom that decays it would be possible in principle to check the presence of vacuum fluctuations and their role in the decay process. Of course this possibility is valid only if the vacuum fluctuations are not too frequent, i.e. the occupation number fluctuations is not too small to be detected. In any case it is essential that in the experiment a large collection of data
is taken to separate the genuine physical effect from the statistical fluctuations.  
Unfortunately it is difficult to have an estimate of how frequent are the vacuum fluctuations.    
\par 
The treatment of spontaneous decay within this model can be easily extended to the case of an initial state (\ref{eq:init}). 
However, it has to be emphasized that the reduction process inherent to the spontaneous decay has a different feature with respect to the one present in the strong measurement process, where the state basis is fixed by the detector \cite{IJQF}, while here the reduction takes place between the ground state and the excited state at the moment of the fluctuation. This is the reason for the normalization of the wave function, as dictated by the Born' s rule, in agreement with the SWF model.  
\par
One could wonder what will happen if a detector is close enough to the atom, so that the reduction of the decaying wave function is dominated by the photon absorption by the apparatus rather than the genuine spontaneous emission of the atom
induced by the vacuum fluctuations. In principle, if the detector is highly sensitive and a response time much shorter than the atomic lifetime, the absorption will occur as soon as the decay channel starts to develop according to the unitary evolution and no exponential law can appear, but rather an almost instantaneous back action of the detector on the atomic state. 
However, if the detector has a response time much longer than the atomic lifetime, the reduction of the decaying wave function
will be mainly produced by the vacuum fluctuations and the exponential law will be recovered.
It has to pointed out that the photon absorption by a detector is initiated by a single elementary process, e.g. the production of an electron in a photo detector, which can produce an electrical signal
through amplification (avalanche). The absorption process is just the time reversal of the emission process, and therefore
it is also triggered by the the nonstandard component of the vacuum fluctuations. This means that the signal will be possibly imprinted by the random occurrence of quantum fluctuations. This cannot occur with a genuine spontaneous emission (distant detector), which will bring only the information that the decay had already occurred, since then the photon and the atomic state are disentangled.

\section{Homodyne detection.}
As it has been emphasized by several authors, the quantum decay of a qubit can have quite different features according to the detection method employed. In particular the quantum jump can be replaced by a finite time process. For a time interval much shorter than its lifetime, an excited atom or qubit evolves according to a unitary deterministic dynamics, as expressed in Eqs. (\ref{eq:U}),(\ref{eq:norm}) and in the previous discussion of the SWF model. During this period an entanglement is established
between the radiation field and the (dipole) state of the qubit. In general the intensity of the radiation is too small to be measured by a device sensitive to the corresponding electric field in a non destructive way, namely without a single photon detection.    
The signal can be amplified by homodyne detection, as we schematically summarize. In this device a beam splitter mixes coherently the radiation field with the local radiation of a laser with the same frequency. If $ \psi $ and $ \alpha $ are the qbit field and the laser field respectively, for a balanced beam splitter in the two output ports one gets the fields   
\begin{eqnarray}
\phi_1 &\,=\,& \frac{1}{\sqrt{2}} ( \alpha + \psi ) \nonumber  \\
\phi_2 &\,=\,& \frac{1}{\sqrt{2}} ( \alpha - \psi )
\label{eq:split}\end{eqnarray}
\noindent
The laser field can be represented by a field operator  $  b  $ and the qubit field by a single quantum mode $ a $, while the laser quantum state will be a coherent state $ | \alpha > $ and  the qubit field will be in a single photon state 
$ | E_{\gamma} > $.  What is measured is the electric current produced by two photo detectors, each one at the two out ports. If one performs the difference $ \Delta i $ between these two currents, it can be shown that, apart constant factor
\beq
\Delta i \,=\, |\alpha| < \psi | ( a e^{\imath \theta} + a^{\dag} e^{-\imath \theta} ) | \psi >
\label{eq:curr}\eeq   
\noindent 
where $ \theta $ is the phase of the laser field. In this way the signal of the qubit field is amplified by the large factor $ |\alpha| $, which is the laser intensity. In Eq. (\ref{eq:curr}) both the time phases $ \exp(\imath\omega t)$ and momentum phases $ \exp(\imath k z)$ drop out because the two components of the field have the same frequency and wavelength. The only time dependence comes from the time dependence of the qubit $ \psi $. The phase $ \theta $ is just the remaining constant phase difference between the two components, which can be adjusted by moving the two branches of the homodyne apparatus.\par 
The mean value appearing in  Eq. (\ref{eq:curr}) is mainly the electric field generated by the qubit at the measuring device,  which is proportional to the dipole of the qubit $ < \sigma_x > $. Here $ \sigma $ is the pseudo-spin operator associated with the qubit. The value of the dipole is determined by the quantum state of the qubit. The differential signal $ d h(t) $ detected by the homodyne
\beq
d h(t) \,\approx\,  < \sigma_x >(t) dt
\label{eq:sig}\eeq         
\noindent marks the evolution of the qubit, in particular from the excited state to the ground state. In sharp contrast with the photon detection, where a quantum jump occurs, in the homodyne detection the evolution of the qubit is in principle smooth and of finite time. The evolution of the qubit is therefore dependent on the quantity that is measured, the number of photons in one case and the electric field in the other case. It has to be stressed that homodyne detection works if the measuring time is much smaller than the lifetime of the excited qubit, as we have assumed, since otherwise a spontaneous decay can occur and the detection is interrupted. \par
However, the smooth signal of Eq. (\ref{eq:sig}) is not the total output that one gets in practice. In fact quantum fluctuations can occur in the homodyne current, due to the possible random photon absorption that generates rapid variation of the homodyne current. Because the photon absorption goes through a very intense sequence of instantaneous quantum jumps, usually these fluctuations are described by an Ito stochastic process $ d W_t $, that has to be added to the smooth current of Eq. (\ref{eq:sig}), to get the total signal, to be compared with the one experimentally observed. Of course the stochastic process is infinitely rapid, and one observes simply an irregular signal of finite time fluctuations. Besides the additional homodyne signal, the photon detections produce a back action on the quantum state of the qubit. In fact the emitted field and the state of the qubit are entangled during the unitary evolution, and the photon absorption process perturbs necessarily the wave function of the qubit. To see this one can notice that the photon detection can be described by the action of the photon annihilation operator $ \hat{A} $ on the the quantum state. In agreement with Eq. (\ref{eq:split}), one gets  
\beq
 \hat{A} | \alpha > | \phi_q > \,=\, ( b \,+\, a ) | \alpha > | \phi_q > \,=\, ( \alpha \,+\, a ) | \alpha > | \phi_q > 
\label{eq:A}\eeq  
\noindent where $ | \phi_q > $ is the state of the qubit. In terms of the excited state $ | e > $ and the ground state 
$ | g > $ it can be expanded as
\beq
| \phi_q > \,=\, C_e | e > \,+\, C_g | g > \,+\, C_2 | g > | E_{\gamma} >
\label{eq:qbit} \eeq
\noindent To first order in the small parameter $ r = C_g/\alpha $, one can verify that the annihilation of Eq. (\ref{eq:A}), taking into account the reormalization of the wave function, results in the state
\beq
| \Psi' > \,=\, \Big[\, | \phi_q > - r C_g\, (\, C_e |e > + C_2 | g > | E_{\gamma} >\, )\, + r (\, 1 - C_g^2\, ) |g >\, \Big] | \alpha >
\label{eq:AA}\eeq
\noindent This result indicates that a fraction of the order $ r $ has been added to the ground state amplitude, and correspondingly a similar fraction has been subtracted from the other two components. This is the back reaction of a single photon detection on the qubit state. The effect is very small, since the laser intensity $ \alpha $ is very large, but also the number of photon detections is very large, which results, as already mentioned, in a perturbation of the qubit state that can be idealized by an Ito stochastic process that acts on the qubit state by an annihilation photon operator with an infinitesimal strength. The perturbation of the qubit density matrix $ \rho $ by this stochastic back action can the be written
\beq
 \delta \rho \,=\, dW_t \Big( a \rho + \rho a^\dag - Tr( a \rho + \rho a^\dag ) \rho \Big)
\label{eq:ro}\eeq     
\noindent where the second term is subtracted to make traceless the perturbation, which corresponds to the renormalization of the qubit wave function. The factor $ dW_t $ is the Ito stochastic process, in agreement with the previous discussion.
\par 
Notice that, according to Eq. (\ref{eq:AA}), to the extent that the laser field is a perfect coherent state, the absorption quantum jump occurs on the qubit field. This establishes a direct link between the fluctuation of the homodyne current and the back action on the qubit state. The Ito stochastic process that is assumed to rule this link is of course an idealization
for the set of absorption processes that occur in both photon detectors. Through amplification, each absorption gives a small pulse to the homodyne current, in one direction or in the other. For a perfect balanced homodyne apparatus these pulses have zero mean value, and the homodyne detection is in reality an irregular sequence of small current jumps. Besides that, in practice the signal is smoothed by the finite time resolution of the current measurement, which cannot resolve each individual jump. However, it would be possible to analyze statistically the fluctuating part of the homodyne signal $ \Delta i $ to reveal its possible hidden structure. In particular, the simplest quantity to look a is the auto-correlation $ \zeta(\tau) $ of the observed signal  
\beq
\zeta(\tau) \,=\, < \Delta(t) \Delta(t + \tau) > 
\label{eq:zeta}\eeq        
\noindent where the average is over repeated runs of the observations. For a long sample of signal recording the function should be independent of the initial time $ t $. If the underlying random process is close enough to a white noise, the function will be sharply peaked at $ \tau = 0 $. Deviations from this behavior would indicate correlations present in the signal. In the SWF and QMOP models this is not expected to occur, because the reduction corresponding to a single photon absorption is considered the result of a continuous monitoring of the photon state, where no correlation is present.     
In the NSM model the absorption process is triggered by the vacuum fluctuations, which will introduce rapid fluctuations in $ \zeta $ as a function of $ \tau $. In particular, if the vacuum fluctuations are interspersed by periods of no fluctuation, the correlation function will present irregular dips. The signal fluctuations can be further analyzed by the Fourier transform of $ \zeta(\tau) $. Of course this irregular fluctuations could be too rapid to be detected by the homodyne apparatus, but in principle they must be present.   
\par
In ref. \cite{Naghi} the observed homodyne current was recorded and used calculate the corresponding back action on the evolution of an artificial qubit. It was show that the back action produce a random evolution of the qubit, in particular the excited state moves toward the ground state by a stochastic finite time diffusion process instead of a quantum jump. This
clearly demonstrates the relevance of the method of detection of the photon field.    
\section{Fluorescence emission.}
If an atom is under the action of an external periodic field, it can undergo oscillations between two states, with a frequency that is proportional to the strength of the field. In particular if the frequency of the field equals the one corresponding to the transition between the ground state and an excited state, the population $ n(t) $ of the excited state is oscillating according to
\beq
n(t) \,=\, \sin^2(\Omega_R t )
\label{eq:occ}\eeq  
\noindent provided $ n(0) = 0 $. The frequency $ \Omega_R $ is given by 
\beq
\Omega_R \,=\, e \mathbf{\mu} \cdot \mathbf{E_0} 
\label{eq:OR}\eeq
\noindent where $ \mathbf{\mu} $ is the dipole matrix element of the transition and $ \mathbf{E_0} $ is the strength of the external field. The oscillations are usually referred to as "Rabi oscillations" and $ \Omega_R $ the Rabi frequency \cite{Rabi}. Eq. (\ref{eq:OR}) neglects the spontaneous emission and therefore it is valid for time much shorter than the lifetime of the excited state. For longer time spontaneous emissions will occur randomly during the oscillations. In a statistically description of the dynamics one considers the equation of motion for density matrix $ \rho $ and the spontaneous emission acts as a friction of the oscillations, which are then damped. If one neglects other sources of decoherence, which depend on the particular system considered, the equation of motion for the average occupation number of the excited state can be written\cite{Torrey,Allen}
\beq
n(t) \,=\, \frac{1}{2} \Big[ 1 \,-\, \big( \cos(\Omega_R t) + \frac{\gamma}{2\Omega_R} \sin(\Omega_R t) \big) 
e^{-\Gamma t/2} \Big]
\label{eq:Torrey}\eeq
\noindent where $\Gamma$ is the decay rate. One recognizes a damped oscillation regime. For large time the occupation number of the excited state, as well as of the ground state, tends to $ 1/2 $, i.e. the atom has equal probability to be in the excited and in the ground states. This is just the temporal average of Eq. (\ref{eq:occ}).
\par One way to observe the damped Rabi oscillations of Eq. (\ref{eq:Torrey}) is by means of fluorescence in continuous-wave field experiments. In fact the probability of spontaneous emission is proportional to the occupation number of the excited state, so the fluorescence intensity will follow the trend of the occupation number. This method was followed in ref. \cite{QD},
where the onset of (damped) Rabi oscillations was observed in a quantum dot as the intensity of the driving (laser) field was increased.\par
In the QMOP and STW models the fluorescence intensity is expected to be a smooth function of time, with fluctuations due only to the statistical uncertainty and possible other noises produced by the detectors. In the NSM model the fluctuations in the signal should include an underlying intrinsic random fluctuation in the dynamics, as in the case of a single isolated atom which spontaneously decays discussed above. A genuine statistical fluctuation is expected to be gaussian and symmetric with respect to the average. In the NSM model this is not the case, because the fluorescence intensity is the sum of all photon emissions (per unit of time) as quantum jumps generated by nonstandard vacuum fluctuations and therefore its intensity fluctuations are expected to follow the law of Eq. (\ref{eq:Da}). They could be observed, provided they overcome the statistical fluctuations. The latter can be reduced by increasing the number of runs of the experiment.
This genuine physical effect is larger for faster decay rate, as it can be seen from Eqs. (\ref{eq:aa}). Of course the possibility of detecting this random part of the dynamics is limited by the time resolution of the detector, and by the frequency of the vacuum fluctuations, which unfortunately is unknown.  
\section{Summary and conclusions.}
This paper addresses one of the fundamental problem in Quantum Mechanics, the dynamics of a single atom or qubit which undergoes spontaneous emission from an excited state. The process itself of photon emission is expected to be instantaneous, but the dynamics before the emission depends on the theoretical model adopted. In fact the emission is in general stochastic and for its description one needs to make assumptions about the features and possible origin of this random process. The work compares three different theoretical models. It is shown that the 'open quantum system' model and the 'stochastic wave function' model are essentially equivalent. They both assume that the stochastic character of the spontaneous emission has to be considered a basic physical law, and therefore no explanation is needed. Ultimately they provide similar descriptions.  
The third model is part of a realistic extension of Quantum Mechanics proposed recently, where most of the open questions,
in particular the measurement problem, find a solution. In this model spontaneous emission is triggered by the vacuum fluctuations, that include a nonstandard component and are considered the origin of the stochasticity. In this case there is no 'continuous' monitoring of the system, but rather an irregular field which acts on the system. It turns out that the spontaneous emission must contain an irreducible fluctuation not present in the other models. It is then proposed to analyze in a different prospect three types of experiments, already realized in several laboratories. It is discussed to what extent these new analysis could reveal the presence of the fluctuations. This would also possibly detect the vacuum fluctuations, which in standard Quantum Mechanics are considered responsible of the Lamb' s shift and Casimir effect. Then the nonstandard component of the vacuum fluctuations would be identified as the origin of the spontaneous emission.

\end{document}